\documentstyle[preprint,aps,epsfig]{revtex}
\begin{document}
%\draft
%@twocolumnfalse\endcsname

\title{Lorentz transformation and vector field flows}
\author{Shao-Hsuan Chiu$^{1}$\thanks{Email address: schiu@frostburg.edu}
       and T. K. Kuo$^{2}$\thanks{Email address: tkkuo@physics.purdue.edu}}
\address{$^{1}$Department of Physics, Frostburg State University, Frostburg,
   MD 21532, USA}
\address{$^{2}$Department of Physics,
     Purdue University, West Lafayette, IN 47907, USA}
\maketitle
\newif\iftightenlines\tightenlinesfalse
\tightenlines\tightenlinestrue

\begin{abstract}

The parameter changes resulting from a combination of
Lorentz transformation are shown to form
vector field flows.  The exact, finite Thomas rotation
angle is determined and interpreted intuitively.
Using phase portraits, the parameters evolution can be
clearly visualized.  In addition to identifying the fixed
points, we obtain an analytic invariant, which correlates
the evolution of parameters.

\end{abstract}
%\pacs{PACS numbers: }
%\vskip2pc]
%\pagenumbering{arabic}

\newpage

\section{Introduction}

One of the basic questions in the Lorentz transformation is velocity addition.
Although algebraic formulas exist~\cite{jackson}, the velocity transformations
are quite complicated owing to their non-commutative nature.
The conceptual complexity arises mainly from the counterintuitive
consequences of the Thomas rotation.  Furthermore,
the determination of the transformation parameters is in general
quite involved.

Interestingly, it has recently
been shown~\cite{kwc} that the two-flavor neutrino mass matrix in
the seesaw model~\cite{seesaw} exhibits a Lorentz group-like structure,
from which the constraint on
the mixing angle and the hierarchical structure of the neutrino
masses can be established.  On the other hand,
the RGE (Renormalization Group Equation) running of the
neutrino mass and mixing angle between high and low
energy scales can be illustrated as the flow of vector field~\cite{kpw}.
In the literature, however, the link between
Lorentz transformation and vector field flows
has not been investigated even though there seems an inherent
connection between them.

Following a general and intuitive approach, we shall analyze the
flow-like structure of the Lorentz velocity transformation,
and show that some intriguing
results can be obtained directly along this line.
Starting from the simple commutation relations of the $2 \times 2$ spinor
algebra, we first construct the Lorentz velocity transformation
and obtain the exact, finite Thomas rotation angle associated with
the transformation.
This approach leads directly to an alternative physical
interpretation of the Thomas rotation angle.
We then derive and solve the differential equations
for the transformation parameters and illustrate their properties in the
phase portraits.
In particular, an invariant governing the evolution of velocity and
direction in the Lorentz velocity addition is established from
our results.
As an example, the collimation effect of
the relativistic decay is interpreted as the approach to
a fixed point in a vector field flow problem.

\section{Lorentz velocity transformation}

The order of successive Lorentz transformations plays a
crucial role when two inertial reference frames are related.
This property is indicated by the commutation relations~\cite{ryder}:

%%%%%%%%%%%%%%%%%%%%%%%%eq(1)%%%%%%%%%%%%%

\begin{eqnarray}
 & [J_{i},J_{j}] & = i\epsilon_{ijk} J_{k}, \nonumber \\
 & [J_{i},K_{j}] & =  i\epsilon_{ijk} K_{k}, \nonumber \\
 & [K_{i},K_{j}] & =  -i\epsilon_{ijk} J_{k},
\end{eqnarray}
%%%%%%%%%%%%%%%%%%%%%%%%%%%%%
where $J_{i,j,k}$ and $K_{i,j,k}$ are the infinitesimal
generators of rotations and pure Lorentz boosts, respectively.
The Thomas precession is known to
originate from this non-commutability of the generators:
A new reference frame reached by two
successive Lorentz boosts cannot be reached by a third,
pure boost from the original frame without a Thomas rotation.
In the literature, the infinitesimal Thomas rotation
angle is usually calculated from a continuous application of infinitesimal
Lorentz transformations consisting of rotations and
boosts~\cite{jackson},
while the finite Thomas rotation angle can be determined by
a variety of approaches~\cite{mc,hu,aau}.
In this article we employ
the simple properties of Pauli matrices instead,
for the determination of finite Thomas rotation angle
as well as other parameters in the Lorentz velocity transformations.

It is well known that
$J_{i}=\sigma_{i}/2$ (rotations) and $K_{i}=i\sigma_{i}/2$ (boosts)
are two-dimensional representations of the Lorentz group:
A finite rotation about an arbitrary axis $\hat{n}$ through an angle
$\Theta$ is written as
$\mathbf{R}$ = $\exp(\frac{i\Theta \vec{\sigma} \cdot \hat{n}}{2})$,
while $\mathbf{B}$ = $\exp(\frac{-\kappa\vec{\sigma} \cdot \hat{n}}{2})$
represents a pure boost along an arbitrary direction $\hat{n}$, with
$\kappa$ the rapidity parameter.
Without loss of generality, we consider the addition of two pure boosts
by choosing one boost of rapidity parameter $\eta$ along the
direction $\hat{n}_{\theta_{0}}=(\sin \theta_{0} \hat{x}+
\cos \theta_{0} \hat{z})$:
%%%%%%%%%%%%%%%%%%%%%%%%%%%%%eq(3)
\begin{equation}
	\mbox{\boldmath $\beta_{1}$} = \tanh \eta (\sin \theta_{0} \hat{x}+
                       \cos \theta_{0} \hat{z}),
\end{equation}
%%%%%%%%%%%%%%%%%%%%%%%%%%%%%%%%%%%%%
and the other of rapidity $\xi$ along $\hat{z}$:
%%%%%%%%%%%%%%%%%%%%%%%%%%%%%%%%%eq(4)
\begin{equation}
	\mbox{\boldmath $\beta_{2}$} = \tanh \xi \hat{z},
\end{equation}
%%%%%%%%%%%%%%%%%%%%%%%%%%%%%%%%%%%%%%%%%%%
as shown in Fig.1.  The combination of the two pure boosts
%%%%%%%%%%%%%%%%%%%%%%%%%%%%%%%%%%%%%%%%eq(5)
\begin{equation}
       L  =  e^{-\frac{\xi}{2} \vec{\sigma} \cdot \hat{z}}
             e^{-\frac{\eta}{2} \vec{\sigma} \cdot \hat{n}_{\theta_{0}}}
          =   e^{-\frac{\xi}{2} \sigma_{3}}
              e^{-\frac{\eta}{2}
              (\sigma_{3} \cos\theta_{0}+\sigma_{1}\sin \theta_{0})}
\end{equation}
%%%%%%%%%%%%%%%%%%%%%%%%%%%%%%%%%%%%%%%%%%%%%%%%%%%%%%%%%%%%
is equivalent to a third boost plus a rotation about an axis
parallel to $\hat{y}$:
%%%%%%%%%%%%%%%%%%%%%%%%%%%%%%%%%%%%%eq(6)
\begin{eqnarray}
L & = & e^{-\frac{\lambda}{2}(\sigma_{3}\cos \theta +\sigma_{1} \sin \theta)}
   e^{i\frac{\tau}{2} \sigma_{2}} \nonumber \\
  & = & e^{-i\frac{\theta}{2} \sigma_{2}} e^{-\frac{\lambda}{2} \sigma_{3}}
  e^{i\frac{\theta +\tau}{2} \sigma_{2}},
\end{eqnarray}
%%%%%%%%%%%%%%%%%%%%%%%%%%%%%%%%%%%%%%%%%%%%%%%%%%%%%%%
where $\tau$ is the Thomas rotation angle
and $\lambda$ represents the third rapidity in the direction
$\hat{n}_{\theta}=(\sin \theta \hat{x}+\cos \theta \hat{z})$, with
$0 \leq \theta \leq \pi$.
Given two boosts of parameters $\xi$ and $\eta$, which are separated by
an angle $\theta_{0}$,
the new parameters associated with the third boost,
$\lambda$ and $\theta$, can be derived
from the simple product $LL^{T}$~\cite{kwc}:
%%%%%%%%%%%%%%%%%%%%%%%%%%%%%%%%%%%%%%%%%
\begin{equation}
   \tan \theta=
\frac{\sin \theta_{0} \sinh \eta}{\sinh \xi \cosh \eta+
                \cos \theta_{0} \cosh \xi \sinh \eta},
\end{equation}
%%%%%%%%%%%%%%%%%%%%%%%%%%%%%%%%%%%%%%%%%%%%%
%%%%%%%%%%%%%%%%%%%%%%%%%%%%%%%%%%%%%%%%%%%%%%%%%eq(10)
\begin{equation}
 \cosh \lambda=
\cosh \xi \cosh \eta + \cos \theta_{0} \sinh \xi \sinh \eta.
\end{equation}
%%%%%%%%%%%%%%%%%%%%%%%%%%%%%%%%%%%%%%%%%%%%%%
The above two equations are equivalent to eq.(11.32)
in Jackson~\cite{jackson}.

\section{The Thomas rotation angle}

To determine the finite Thomas rotation angle $\tau$, we now
consider $L^{T}L$.  From eq.(4),
%%%%%%%%%%%%%%%%%%%%%%%%%%%%%%%%%%%%%%%%eq(11)
\begin{equation}
L^{T}L=e^{-\frac{\eta}{2}(\sigma_{3}\cos \theta_{0}+
        \sigma_{1}\sin \theta_{0})}
      e^{-\xi \sigma_{3}}
     e^{-\frac{\eta}{2}(\sigma_{3}\cos \theta_{0}+
        \sigma_{3}\sin \theta_{0})},
\end{equation}
%%%%%%%%%%%%%%%%%%%%%%%%%%%%%%%%%%%%%%%%%%%%%%%%
while eq.(5) gives
%%%%%%%%%%%%%%%%%%%%%%%%%%%%%%%%%%%%%%%%%%eq(12)
\begin{equation}
L^{T}L=e^{-i\frac{\theta +\tau}{2} \sigma_{2}}
      e^{-\lambda \sigma_{3}}
     e^{i\frac{\theta +\tau}{2} \sigma_{2}}.
\end{equation}
%%%%%%%%%%%%%%%%%%%%%%%%%%%%%%%%%%%%%%%%%%%%%%%%%%%
Eq.(8) can further be simplified by using the identity
%%%%%%%%%%%%%%%%%%%%%%%%%%%%%%%%%%%%%%%%%%%%%%eq(13)
\begin{equation}
e^{-i\frac{\theta_{0}}{2} \sigma_{2}}e^{-\frac{\eta}{2} \sigma_{3}}
e^{i\frac{\theta_{0}}{2}\sigma_{2}}=
e^{-\frac{\eta}{2} (\sigma_{3}\cos \theta_{0}+\sigma_{1}\sin \theta_{0})},
\end{equation}
%%%%%%%%%%%%%%%%%%%%%%%%%%%%%%%%%%%%%%%%%%%%%%%%%%%%%%%%%%%%%%%%%
which leads to
%%%%%%%%%%%%%%%%%%%%%%%%%%%%%%%%%%%%%%%%%%%%%%%%%%%%eq(14)
\begin{equation}
L^{T}L=e^{-i\frac{\theta_{0}}{2} \sigma_{2}}(e^{-\frac{\eta}{2} \sigma_{3}}
      e^{-\xi \sigma_{(-\theta_{0})}}
             e^{-\frac{\eta}{2} \sigma_{3}})e^{i\frac{\theta_{0}}{2}
              \sigma_{2}},
\end{equation}
%%%%%%%%%%%%%%%%%%%%%%%%%%%%%%%%%%%%%%%%%%%%%%%%%%%%%%%%%%%%
where $\sigma_{(-\theta_{0})} \equiv \sigma_{3}\cos (-\theta_{0})+
       \sigma_{1}\sin (-\theta_{0})$.
Eqs.(9) and (11) lead to a simple relation:
%%%%%%%%%%%%%%%%%%%%%%%%%%%%%%%%%%%%%%%%%%%%%%%%%%%%%%%%%eq(15)
\begin{equation}
\tan (\theta + \tau -\theta_{0}) \equiv \tan \phi=
\frac{-\sin \theta_{0} \sinh \xi}{\cosh \xi \sinh \eta
                +\cos \theta_{0} \sinh \xi \cosh \eta}.
\end{equation}
%%%%%%%%%%%%%%%%%%%%%%%%%%%%%%%%%%%%%%%%%%%%%%%%%%%%%
Here, $\phi$ specifies the direction of the resultant third boost
when $\mbox{\boldmath $\beta_{1}$}$ and $\mbox{\boldmath $\beta_{2}$}$
are applied in a reverse order.
The finite Thomas rotation angle is then given by
%%%%%%%%%%%%%%%%%%%%%%%%%%%%%%%%%%%%%%%%%%%%%%%%%%%%%eq(17)
\begin{equation}
\tan \tau =
  \frac{\tan \phi+\tan(\theta_{0}-\theta)}
  {1-\tan\phi \tan(\theta_{0}-\theta)}.
\end{equation}
%%%%%%%%%%%%%%%%%%%%%%%%%%%%%%%%%%%%%%%%%%%%%%%%%%%%%%%%%%%%%%%%

Eqs.(6), (7), and (13) are exact, and valid for the finite
Lorentz transformations.  In the limit of
infinitesimal Lorentz transformation,
we may consider a finite boost $\beta_{1} = \tanh \eta$, followed by an
infinitesimal boost $\beta_{2}=\tanh(\Delta \xi) \cong \Delta \xi$.
The infinitesimal Thomas rotation angle then becomes
%%%%%%%%%%%%%%%%%%%%%%%%%%%%%%%%%%%%%%%%%%%eq(20)
\begin{equation}
\Delta\tau \cong \frac{\sin\theta_{0}}{\sinh\eta}(\cosh\eta - 1)\Delta\xi,
\end{equation}
%%%%%%%%%%%%%%%%%%%%%%%%%%%%%%%%%%%%%%%%%%%%%%%%
which agrees with the result in Section 11.8 of Ref.~\cite{jackson}.
Note that our notations are slightly different from that of
Ref.~\cite{jackson}: our $\mbox{\boldmath $\beta_{1}$}$ and
$\mbox{\boldmath $\beta_{2}$}$ correspond to $\mbox{\boldmath $\beta$}$
and $\gamma^{2} \delta \mbox{\boldmath $\beta_{\parallel}$} +
\gamma \delta \mbox{\boldmath $\beta_{\perp}$}$ in Ref.~\cite{jackson},
respectively.
We also note that eq.(13) agrees with eq.(37)
of Ref.~\cite{aau} up to an overall sign, which is due merely
to the different sense of rotation.

An alternative physical interpretation of the Thomas rotation angle becomes
clear from our formulation as we examine
two boosts combined in reverse orders.
From eq.(4) and eq.(5),
%%%%%%%%%%%%%%%%%%%%%%%%%%%%%%%%%%%%%%%%%%%%%eq(24)
\begin{equation}
e^{-\frac{\xi}{2} \sigma_{3}}e^{-\frac{\eta}{2} \sigma_{\theta_{0}}}
=e^{-\frac{\lambda}{2} \sigma_{\theta}}e^{i\frac{\tau}{2} \sigma_{2}},
\end{equation}
%%%%%%%%%%%%%%%%%%%%%%%%%%%%%%%%%%%%%%%%%%%%%
where $\sigma_{\theta_{0}} \equiv \sigma_{3}\cos \theta_{0}+
     \sigma_{1}\sin \theta_{0}$, and
 $\sigma_{\theta} \equiv \sigma_{3}\cos \theta+
 \sigma_{1}\sin \theta$.
It follows that
%%%%%%%%%%%%%%%%%%%%%%%%%%%%%%%%%%%%%%%%eq(25)
\begin{equation}
e^{-\frac{\xi}{2} \sigma_{3}}e^{-\frac{\eta}{2} \sigma_{\theta_{0}}}
       e^{-i\frac{\tau}{2} \sigma_{2}}
=e^{-\frac{\lambda}{2} \sigma_{\theta}}.
\end{equation}
%%%%%%%%%%%%%%%%%%%%%%%%%%%%%%%%%%%%%%%%%%%%%%
Since $\sigma_{\theta}$ is symmetric, it follows that
\begin{equation}
e^{i\frac{\tau}{2} \sigma_{2}}(e^{-\frac{\eta}{2} \sigma_{\theta_{0}}}
e^{-\frac{\xi}{2} \sigma_{3}})
e^{i\frac{\tau}{2} \sigma_{2}}=e^{-\frac{\xi}{2} \sigma_{3}}
e^{-\frac{\eta}{2} \sigma_{\theta_{0}}}.
\end{equation}
%%%%%%%%%%%%%%%%%%%%%%%%%%%%%%%%%%%%%%%%%%%%%%%%%%%%%%%

Eq.(17) implies that the combination of two boosts,
$\exp(-\frac{\xi}{2} \sigma_{3})
\exp(-\frac{\eta}{2} \sigma_{\theta_{0}})$, is related to its reverse,
$\exp(-\frac{\eta}{2} \sigma_{\theta_{0}})\exp(-\frac{\xi}{2} \sigma_{3})$,
by two identical rotations, $e^{i\frac{\tau}{2} \sigma_{2}}$.
In other words, operating a rotation on a reference frame
before and after two successive boosts  would bring
this frame to the same reference
frame that is reached by the same two boosts operated in reverse order.
The angle associates with this particular rotation is the
Thomas rotation angle, whose existence is thus directly related to
the non-commutativity of the Lorentz transformations.

\section{Lorentz transformations and vector field flows}

A geometrical visualization of the Lorentz transformation can
be realized as a vector field flow problem.  Let us start from
any Lorentz transformation specified by the parameters
$(\theta,\beta,\tau)$, which we denote as $L_{(\theta,\beta,\tau)}$.
A second infinitesimal boost would change $L$ according to
$L_{d\xi} L_{(\theta,\beta,\tau)} =
L_{(\theta+d\theta,\beta+d\beta,\tau+d\tau)}$.
Clearly, this change can be represented as an
infinitesimal vector in a 3D space originating from
the $(\theta,\beta,\tau)$ point.  The general problem for
arbitrary $\xi$ and $(\theta,\beta,\tau)$ is thus succinctly
described by a vector field flow problem.

To quantify the vector flows we need three differential equations for
$\theta$, $\beta$ ($\beta \equiv \tanh \lambda$), and $\tau$ which can
be derived from eqs.(6), (7), and (13):
%%%%%%%%%%%%%%%%%%%%%%%%%%%%%%%%%%%%%%%%%%%%%%%%%%%%%%%eq(28)
\begin{equation}
\frac{d\theta}{d\xi}=\frac{-\sin\theta}{\beta},
\end{equation}
%%%%%%%%%%%%%%%%%%%%%%%%%%%%%%%%%%%%%%%%%%%%%%%%%%%%%%%%
%%%%%%%%%%%%%%%%%%%%%%%%%%%%%%%%%%%%%%%%%%%%%%%%%%%%%eq(29)
\begin{equation}
\frac{d\beta}{d\xi}=\cos\theta (1-\beta^2),
\end{equation}
%%%%%%%%%%%%%%%%%%%%%%%%%%%%%%%%%%%%%%%%%%%%%%%%%%%%%
%%%%%%%%%%%%%%%%%%%%%%%%%%%%%%%%%%%%%%%%%%%%%%%%%eq(30)
\begin{equation}
\frac{d\tau}{d\xi} = \frac{\sin\theta}{\beta} (1-\sqrt{1-\beta^2}).
\end{equation}
%%%%%%%%%%%%%%%%%%%%%%%%%%%%%%%%%%%%%%%%%%%%%%%%%%%%%%%%%%%

We first examine the relation between the two angles, $\theta$ and $\tau$.
From eq.(20)
we see that at small $|\mbox{\boldmath $\beta$}|$, the Thomas rotation angle $\tau$
does not vary significantly with $\xi$:
$\frac{d\tau}{d\xi} \approx 0$, while
$\theta$ varies rapidly according to eq.(18).
On the other hand, $\frac{d\theta}{d\xi}$ slows down and
$\frac{d\tau}{d\xi}$ speeds up as $|\mbox{\boldmath $\beta$}|$ increases.
From eqs.(18) and (20) we have
%%%%%%%%%%%%%%%%%%%%%%%%%%%%%%%%%%%%%%eq31
\begin{equation}
\frac{d\tau}{d\theta} = -(1-\sqrt{1-\beta^{2}}).
\end{equation}
%%%%%%%%%%%%%%%%%%%%%%%%%%%
The solution, given by
%%%%%%%%%%%%%%%%%%%%%%%%%%%%%%%%%%%%%%%%%%%eq(32)
\begin{equation}
\tau - \tau_{0} = \frac{1-\cosh\lambda}{\cosh\lambda}
(\theta - \theta_{0}),
\end{equation}
%%%%%%%%%%%%%%%%%%%%%%%%%%%%%%%%%%%%%%%%%%%%%
clearly describes the variation of $\tau$ and $\theta$ for a
given boost $\beta$.
For a given $|\mbox{\boldmath $\beta$}|$,
eq.(18) implies that the rate of change of
$\theta$ is maximum at $\theta = \pi/2$,
where the two boosts are perpendicular.
In addition, eqs.(12) and (13)
imply that if $\beta_{1} \approx 1$ and $\beta_{2} \approx 1$,
then $\tan \phi \cong 0$ and
$\tau \cong \theta_{0}$.  Therefore, when the two successive
boosts are each close to light speed, the value of Thomas rotation angle
associated with the transformation approaches that of the
angle between the two boosts.  Furthermore, $\tau = 0$ if the
two successive boosts are co-linear.

As for $\tau$ and $\lambda$, eqs.(19) and (20) give rise to the
equation
%%%%%%%%%%%%%%%%%%%%%%%%%%%%%%%%%%%%%
\begin{equation}
\frac{d\tau}{d\beta} = \frac{\tan\theta (1-\sqrt{1-\beta^{2}})}
{\beta (1-\beta^{2})},
\end{equation}
%%%%%%%%%%%%%%%%%%%%%%%%%%%%
with the solution given by
%%%%%%%%%%%%%%%%%%%%%%%%%%%%%%%%%%%%%%%%%eq33%%%%%%%
\begin{equation}
\tan\theta = \frac{\tau-\tau_{0}}{\ln(\cosh \lambda -1)-
\ln(\cosh \lambda_{0} -1)},
\end{equation}
which relates the change of Thomas angle to the boost parameters
for a given angle $\theta$.

Finally, from eqs.(18) and (19) we have
%%%%%%%%%%%%%%%%%%%%%%%%%%%%%%%%%%%%%%%%%%%%%eq(34)
\begin{equation}
\frac{d\theta}{d\beta} = \frac{-\tan \theta}{\beta (1-\beta^{2})}.
\end{equation}
%%%%%%%%%%%%%%%%%%%%%%%%%%%%%%%%%%%%%%%%%%%%%%%
The solution is given by
%%%%%%%%%%%%%%%%%%%%%%%%%%%%%%%%%%%%%%%%%eq(34)
\begin{equation}
\sin\theta \sinh \lambda = \sin \theta_{0} \sinh \lambda_{0}.
\end{equation}
%%%%%%%%%%%%%%%%%%%%%%%%%%%%%%%%%%%%%%%%%%%%%%
Interestingly, this solution represents an invariant
that correlates the evolution of direction and magnitude of
the resultant boost.  Note that the Thomas rotation angle
$\tau$ does not take part in this invariant.
We also note that in the analysis of neutrino parameters~\cite{kwc},
$\theta$ and $e^{2\lambda}$ represent the mixing angle and
the physical neutrino mass ratio, respectively.

Our results can be illustrated clearly using the phase portraits.
The $\theta-\beta$
phase portrait is shown in Fig.2, where the direction fields are
plotted for increasing $\xi$ and the arrows are
tangent to the trajectories.
There are four fixed points for the evolution equations,
eqs.(18) and (19): $(\beta,\theta)=(1,0)$, $(-1,\pi)$, $(1,\pi)$,
and $(-1,0)$.  The stability~\cite{strogatz} of the fixed points as $\xi$
increases are determined by the eigenvalues of the Jacobian.
Of the four fixed points, $(1,0)$ and $(-1,\pi)$ are stable (attractive)
and the evolution of $\theta$ and $\beta$ is always toward one of the
two points.  On the other hand, the evolution is always directed away from
the unstable (repulsive) fixed points, $(1,\pi)$ and $(-1,0)$.
Note that the two attractive
fixed points are physically identical, so are the two repulsive fixed
points.  This is because $\beta$ of opposite signs simply represent
boosts in opposite directions.
The phase portrait
illustrates the following features:

(I). As $\xi$ increases, starting from any initial condition
$(\beta_{0},\theta_{0})$, the values $(\beta,\theta)$
evolve along trajectories defined by the invariant, eq.(26).
As $\xi \rightarrow \infty$, all of these trajectories converge
onto the stable fixed points
$(\beta,\theta) = (1,0)$ or $(-1,\pi)$,
depending on the signs of the boosts.
Thus, the resultant boost $\mbox{\boldmath $\beta$}$
tends to evolve from its initial value toward
$|\mbox{\boldmath $\beta$}|=1$ (extreme relativistic limit), and
its direction evolves toward that of the second boost
$\mbox{\boldmath $\beta_{2}$}$.  The evolution of direction
also can be understood from eq.(18):
$\frac{d\theta}{d\xi} \geq 0$
for a negative $\beta$ and $\frac{d\theta}{d\xi} \leq 0$
for a positive $\beta$.

(II). The long arrows with slopes approaching infinity
near the region of small
$|\mbox{\boldmath $\beta$}|$ and large
$\sin \theta$ ($\theta \approx \pi/2$) represent the rapid variation of
$\mbox{\boldmath $\beta$}$'s direction in this region.
This property is also implied by eq.(18).

(III). For $|\mbox{\boldmath $\beta$}|=1$,
the slope of the trajectory becomes infinity and
the transformation does not exist.  Physically,
this feature can be interpreted
as no transformation from the rest frame of a
photon to a laboratory frame.

It should be noted that the fixed point structure described
above has a direct realization in the physical process of
relativistic particle decays, e.g.,
$\pi^{0} \rightarrow \gamma \gamma$.
For fast moving $\pi^{0}$ ($\xi \rightarrow \infty$),
most photons (with ($\beta_{0},\theta_{0}$) in the rest frame)
are focused in the forward direction
(with $\beta \rightarrow 1$) in the lab frame.  This can be
clearly visualized in Fig.2, so that the vector field flow
corresponds to the well-known collimation effect of
relativistic decays.  What is not so well-known, however,
is that, during the collimation, there is a simple correlation
between the direction ($\theta$) and the rapidity
($\lambda$), given in eq.(26).

The evolution of $\beta$, $\theta$, and $\tau$ can be visualized in
a 3-D phase portrait.
In Fig.3 the initial conditions corresponding
to $\beta_{2}=0$
and arbitrary $(\beta_{1},\theta_{0})$ lie in the plane
specified by $\tau =0$.  As $\beta_{2}$ increases, $\theta$ and $\beta$
evolve along one of the trajectories in Fig.2
while changing the corresponding
$\tau$ value at varying rates.  This rate of change of $\tau$ is determined
by the local values of $\beta$ and $\theta$ as implied by eq.(20).
If the initial $\beta_{2}$ is chosen to be nonzero,
the initial conditions lie in
a plane specified by a nonzero $\tau$.

\section{Summary}

Simple properties of the spinor algebra provide a general
and effective solution to the problems of
finite Lorentz velocity transformations.  In particular,
we found the exact
solution of the finite Thomas rotation angle.
It has an intuitive physical interpretation, directly related
to the non-commutativity of Lorentz transformations.
Following this line, we then treat the Lorentz velocity transformation as a
vector field flow problem.  In relating the two, we present the
analytical results, eqs.(22), (24) and (26), which come directly from
a set of differential equations, eqs.(18), (19), and (20).  In addition,
we show that the general features of the Lorentz transformation
and the evolution of the parameters can be clearly visualized using
the phase portraits of the parameters.  The attractive fixed point of the
vector field flow describes geometrically the well-known collimation
effect of relativistic decays.  The flow toward the fixed point
follows trajectories given by the invariant, eq.(26).

As a final remark, we note that the invariant, eq.(26),
carries the same form as the general, complex RGE invariant~\cite{kpw}.
However, unlike the running of RGEs for the
general neutrino mass matrix, in which the relative phase of the
mass eigenvalues is nonzero and the evolution depends sensitively
on the choice of initial conditions, the
Lorentz velocity transformation corresponds to a vanishing relative phase
and the evolution of parameters is not as sensitive to the initial
conditions, i.e., physically
there is only one attractive fixed point and one repulsive fixed point.

\acknowledgements
We thank H. Urbantke for pointing out Refs.~\cite{mc} and ~\cite{hu}.
T. K. K. is supported in part by the DOE grant No. DE-FG02-91ER40681.

%%%%%%%%%%%%%%%%%%%%%%%%%%%%%%%%%%%%%%%%%%fig1
\begin{figure}[b]
\epsfig{file=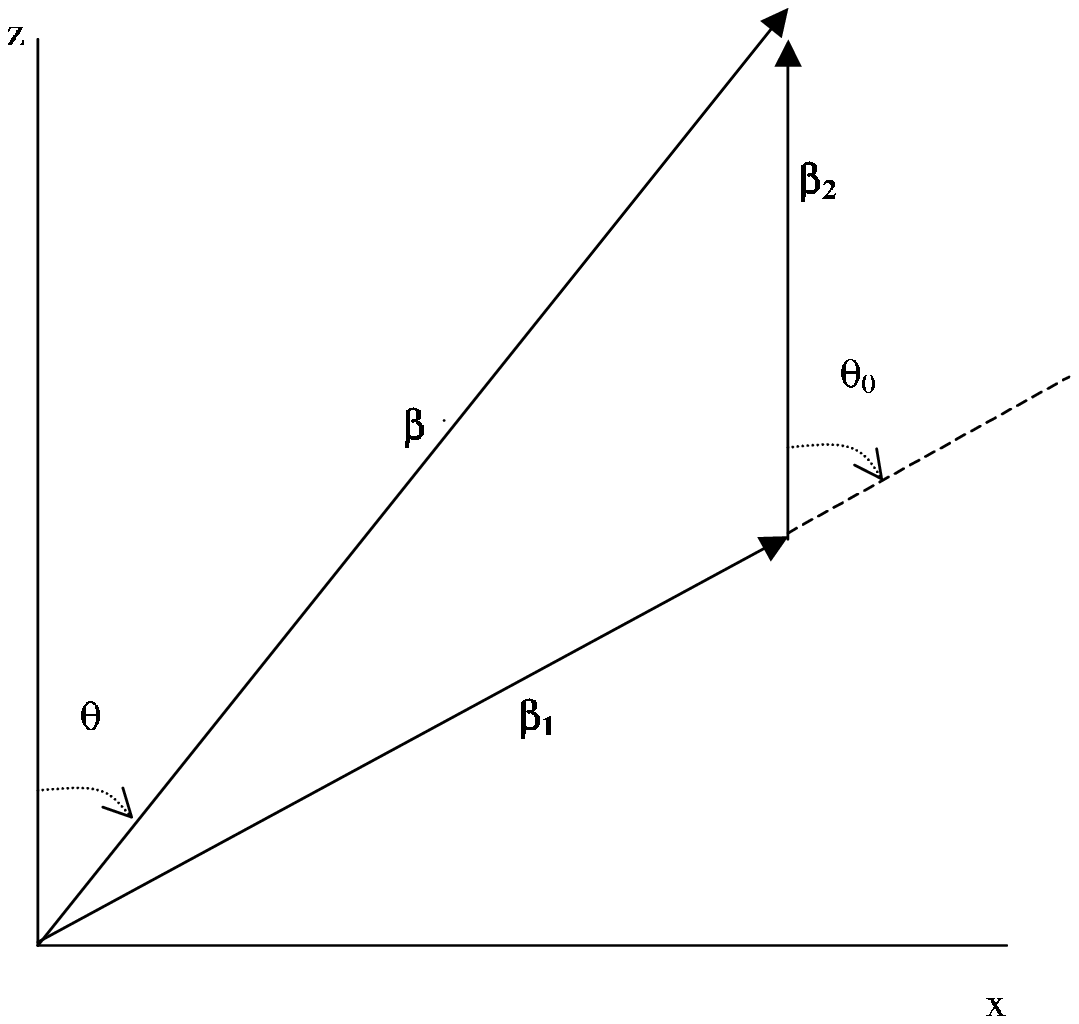,width=13cm}
  \caption{The combination of two boosts, $\mbox{\boldmath $\beta_{1}$}$
  and $\mbox{\boldmath $\beta_{2}$}$, is equivalent to a third boost
  $\mbox{\boldmath $\beta$}$ plus a rotation about $\hat{y}$.
  }
\label{fig1}
\end{figure}

%%%%%%%%%%%%%%%%%%%%%%%%%%%%%%%%%%%%%%%%%%%fig2
\begin{figure}
\epsfig{file=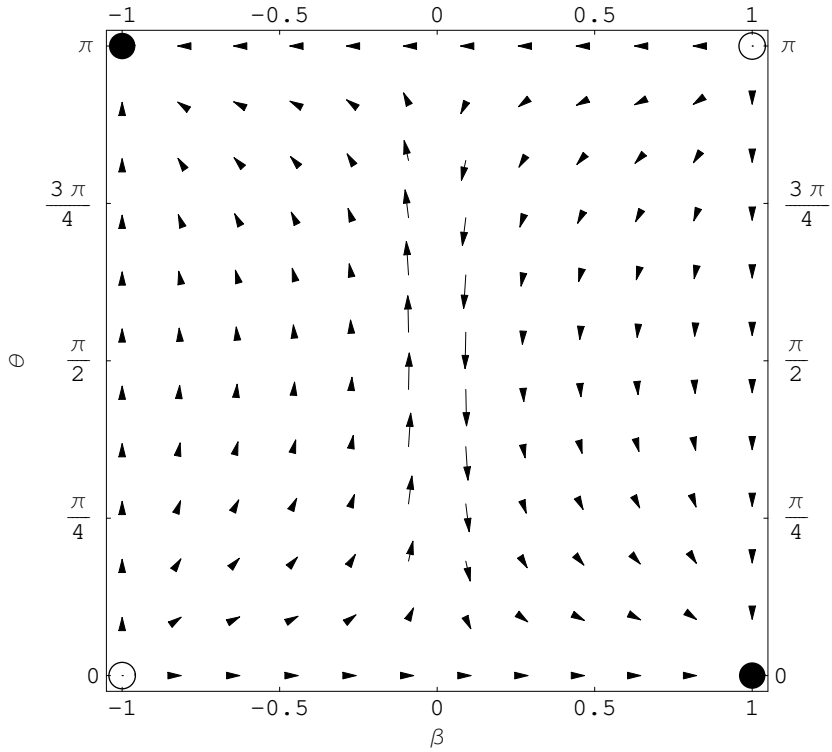,width=12cm}
  \caption{The $\theta - \beta$ phase portrait.  All trajectories evolve
  toward either $(\beta,\theta)=(1,0)$ or $(\beta,\theta)=(-1,\pi)$.
  Note that a negative $\beta$ represents a boost in the direction opposite
  to that of a positive $\beta$.  The trajectories are described
  by eq.(26); the attractive and repulsive fixed points
  are shown in the figure as $\bullet$ and $\circ$, respectively.  Also note that we have used $\beta$ here
  in the graph instead of $\lambda$.
  }
\label{fig2.eps}
\end{figure}
%%%%%%%%%%%%%%%%%%%%%%%%%%%%%%%%%%%%%fig3

\begin{figure}
\epsfig{file=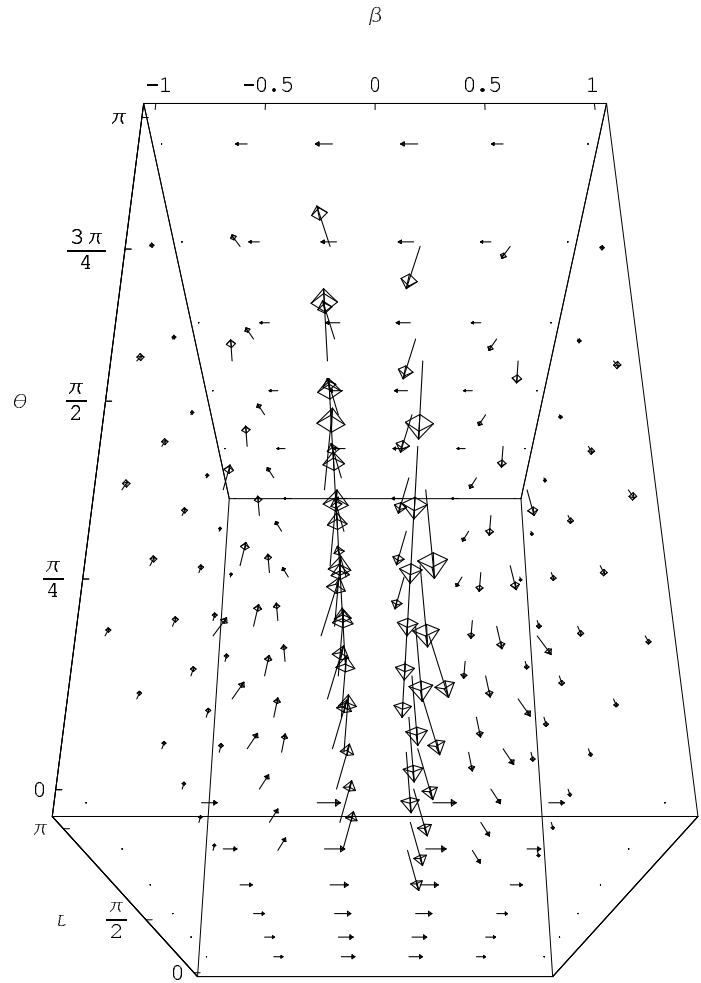,width=12cm}
  \caption{The evolution of $\beta$, $\theta$, and $\tau$.  Note that as
  $\beta \rightarrow 1$, $(\theta,\tau)$ approaches $(0,\pi)$
  for $\beta >0$ and that $(\theta,\tau)$ approaches $(\pi,-\pi)$
   for $\beta < 0$.  Fig.2 corresponds to a ``slice" of this figure
   at a particular $\tau$ value.  The rate of change of $\tau$
   with respect to $\xi$ depends
   on the local values of $\beta$ and $\theta$.
  }
\label{fig3.eps}
\end{figure}

\end{document}